# A phase diagram for bacterial swarming

Avraham Be'er[1,2], Bella Ilkanaiv[1], Renan Gross[3], Daniel B. Kearns[4], Sebastian Heidenreich[5], Markus Bär[5] and Gil Ariel[6]

[1]Zuckerberg Institute for Water Research, The Jacob Blaustein Institutes for Desert Research, Ben-Gurion University of the Negev, Sede Boqer Campus 84990, Midreshet Ben-Gurion, Israel.
[2]Department of Physics, Ben-Gurion University of the Negev 84105, Beer-Sheva, Israel.
[3]Department of Mathematics, Weizmann Institute of Science, Rehovot 76100, Israel.
[4]Department of Biology, Indiana University, Bloomington, Indiana 47405, USA .
[5]Department of Mathematical Modelling and Data Analysis, Physikalisch-Technische Bundesanstalt Braunschweig und Berlin, Abbestrasse 2-12, D-10587 Berlin, Germany .
[6]Department of Mathematics, Bar-Ilan University, Ramat-Gan 52000, Israel.

**Bacterial swarming is a rapid mass-migration, in which thousands of cells spread collectively to colonize a surface. Physically, swarming is a natural example of active particles that use energy to generate motion. Accordingly, understanding the constraints physics imposes on the dynamics is essential to understand the mechanisms underlying the swarming phenomenon. We present new experiments of swarming *Bacillus subtilis* mutants with different aspect ratios and densities. Analyzing the dynamics reveals a rich phase diagram of qualitatively distinct swarming regimes, describing how the shape and density of cells govern the global dynamical characteristics of the entire swarm. Moreover, we show that under standard conditions bacteria inhabit a region of phase space that is associated with rapid mixing and robust dynamics, with homogeneous density and no preferred direction of motion. This contrasts characteristic clustering behavior of self-propelled rods that is recovered only for very elongated mutant species. Thus, bacteria have adapted their physics to optimize the principle functions assumed for swarming.**

Micro-organisms such as bacteria, sperm-, epithelial- and cancer-cells, as well as immune T-cells and even inanimate active particles, generate collective flows and demonstrate a wealth of newly discovered emergent dynamical patterns [Szabó 2006; Sokolov 2007; Ramaswamy 2010; Marchetti 2013; Zhou 2014; Elgeti 2015; Blanch-Mercader 2018; Be'er 2019]. This report addresses the dynamics of swarming bacteria – a biological state to which some bacterial species transition, in which rod-shaped cells, powered by flagellar rotation, migrate rapidly on surfaces *en-mass* [Harshey 2003; Darnton 2010; Kearns 2010; Tuson 2013]. Swarming allows efficient expansion and colonization of new territories, even under harsh and adverse conditions such as starvation or antibiotic stress [Lai 2009; Benisty 2015]. Revealing the biological and physical mechanisms underlying bacterial swarming is therefore a key to our understanding of how bacteria spread and invade new niches.

The transition to swarming involves several critical intra-cellular processes such as an increase in flagellar number and changes in cell shape, suggesting these changes promote favorable swarming conditions [Harshey 2003; Kearns 2010; Tuson 2013; Mukherjee 2015; Be'er 2019]. Quantifying the "quality" of swarming can be done using the tools of statistical physics by analyzing the dynamical properties of large, out-of-equilibrium, self-propelled collectives [Ramaswamy 2010; Vicsek 2012; Bär 2019]. Accordingly, one of the primary goals of such quantification is to obtain a phase diagram that would describe the possible dynamical states of swarms as a function of *controlled* parameters such as density and cell aspect ratio. As bacteria move and grow, they trace an effective path through the phase diagram, transitioning between different swarming regimes. Thus, a phase diagram provides a map that explains how the microscopic mechanical properties of cells, which are regulated by complex bio-chemical cellular processes, govern the global dynamical characteristics of the entire swarm.

The density within a swarm is typically extremely high (up to 0.8 surface coverage), resulting in a combination of short-range steric repulsion and long-range hydrodynamic interactions. Both forces strongly depend on cell aspect ratio and particle density, see e. g., [Peruani 2006; Sokolov 2007; Darnton 2010; Zhang 2010; Peruani 2012, Wensink 2012; Ilkanaiv 2017; Shi 2018, Be'er 2019; Jeckel 2019]. Recently, Jeckel et al [Jeckel 2019] studied expanding colonies of swarming *Bacillus subtilis*, and discovered regions corresponding to single cells, rafts, biofilm and mixed states. However, despite the recent progress in understanding the physics of active matter in general and swarming bacteria in particular, the different pieces do not yet fit together into a comprehensive picture. In particular, it is not clear how the properties of cells determine the dynamical state of a swarm. In order to address this question, in this paper, monolayer swarms of four strains of *B. subtilis* with different aspect ratios – ranging from 5.5 to 19, were analyzed as a function of cell density (Fig. 1A and S1). A custom algorithm enabled tracking of individual cell trajectories, which in turn allowed a comprehensive analysis of both the individual and collective dynamics of bacteria in a swarm. The main results are then expressed in a phase diagram of bacterial swarming (Fig. 1B). The novel two-dimensional set-up of a thin, single-layer of cells, brings out a complex experimentally based phase diagram with various features which could not be obtained with earlier multilayer studies [Zhang 2010, Peruani 2012, Ilkanaiv 2017, Li 2019] or naturally expanding colonies [Jeckel 2019].

In comparison to simulation results for interacting self-propelled rods [Peruani 2006; Ginelli 2010; Wensink 2012; Abkenar 2013; Shi 2018, Jeckel 2019, Bär 2019] and experiments with artificial inanimate systems [Narayan 2007, Kudrolli 2008, Bricard 2013], the swarming phase diagram has several unique characteristics that are not observed in other active systems. These properties may alter, constrain, or even control cells' ability to move collectively in an efficient manner, mix within the colony and spread. As a result, it has direct biological consequences in terms of the ability of bacteria to swarm efficiently.

Deriving a phase diagram requires measuring a large number of collective and individual dynamical statistics for swarms with particular controlled parameters. As described above, we concentrate on cell aspect ratio ($\alpha$) and density ($\rho$) as the fundamental mechanical parameters. Since the typical expansion rate of the colony is about 10 times slower than the typical microscopic swarming speed, different regions of the actively-migrating colony can be considered to be in a quasi-steady state. Therefore, the effect of density can be evaluated by sampling regions that occupy a different number of bacteria per unit area [e.g., Peruani 2012, Jeckel 2019]. Cell shape was manipulated genetically by mutating a few of the robust mechanisms that maintain aspect ratio during growth. Artificially long cells were generated using cells mutated for either MinD or MinJ that control proper medial division in *B. subtilis*. Artificially short cells were generated by overexpression of SwrA, the master activator of flagellar biosynthesis [Guttenplan 2013; Mukherjee 2015]. Overall, four different strains of the same species with various aspect ratios 5.5, 7 (the wild-type), 13 and 19 (Fig. S2) were compared. A wide range of benchmark tests verified that all other motility and expansion-related parameters were the same (including swimming speeds and doubling times in broth, colony expansion rates and surfactant production). See Materials and Methods, SI Text 1 and Fig. S3.

Our analysis identified five dynamical states for bacterial swarms (Fig. 1B). Each state showed distinct dynamical features, as detailed below, which are expressed both at the individual and collective levels. Wild-type bacteria, with a typical aspect ratio of 7 and a wide range of densities (0.2-0.7), inhabited the center of a broad region of the phase space, showing rapid movement and highly efficient spreading. Within this region, the swarming statistics were not sensitive to the density as well as to small changes in the aspect ratio, suggesting that the collective behavior of wild-type swarming cells is particularly robust to fluctuations in density and cell shape. Biologically, this physical robustness of the swarming phase (S) may be advantageous for the colony's survival and expansion, particularly under stress. Collective motion of bacteria changes, however, quite dramatically, if the aspect ratio is increased above a threshold value, that turns out to be around 10 for *Bacillus subtilis* in our study here. For more elongated cells (i.e., larger aspect ratios), cells start to move in separate clusters. A phase with small clusters (SC) composed of a few bacteria is observed if the density is above 0.1 Strikingly, above densities of around 0.25, cells transition into another phase and form large clusters (LC) that can become of the order of the observation window employed here. This transition is reminiscent of behavior of self-propelled rods with short-range alignment interactions (typically due to volume exclusion) in simulations and experiments, see [Bär 2019] and references therein. The observed cluster-size distributions are in line with a kinetic theory describing occurrence of large moving clusters as a specific type of microphase separation characteristic for rod-shaped moving particles like bacteria [Peruani 2006, Peruani 2013]. Independently of the aspect ratio, at low cell densities (<0.1) cells do not move at all – hence we have an immotile phase (IM). At large densities above 0.7 a jamming phase (J) is observed, where cells stop moving due to lack of space. Below, we will characterize in detail the differences between the phases in which cells are moving, namely S, SC, and LC. Already at this stage, we can interpret the phase diagram in Fig. 1B as follows: our main findings show that the behavior at large aspect ratios resembles self-propelled rod dynamics. Hence, we conclude that in the SC and LC phases, collective behavior is dominated by short-range alignment or excluded volume interactions. This is in line with findings for filamentous, very large aspect ratio mutants of *Eschericia coli*, that display behavior dominated by short-range alignment interaction [Nishiguchi 2017]. At small aspect ratios, the collective behavior deviates from the self-propelled rod paradigm. We rationalize this is due to long-range hydrodynamic interactions, which suppress the clustering and density inhomogeneities, qualitatively in line with recent simulation findings [Theers 2019].

In order to better understand the different phases depicted in Fig. 1B, their physical characteristics, as well as the biological implications, we divide the further discussion below along the phase-transition lines. Figure 2A depicts the mean cell speed, which is monotonically increasing with density for each aspect ratio, a hallmark of collective motion – showing that many cells cooperate to produce faster motion. The mean speed is not monotonic in the aspect ratio where wild-type cells seem to be optimal in this regard (see also [Ilkanaiv 2017]). However, speeds depend smoothly on concentration, showing no sharp transitions. In addition, Fig. 2A identifies lower and upper density thresholds, beyond which swarming cannot occur. At very low densities, cells are practically immotile, suggesting a minimal area coverage or number density below which cells cannot move [Kearns 2003]. The minimal speed at which swarming was observed marks the edge of the *immotile* phase IM depicted in Fig. 1B. Note that this phase does not occur in typical active matter systems, including swimming (not swarming) bacteria in bulk or thin films [Sokolov 2007; Sokolov 2012], driven inanimate particles [Narayan 2007; Bricard 2013] or most models of self-propelled particles (SPPs), either discrete or continuous [e.g. Wensink 2012; Abkenar 2013; Ariel 2018], in which isolated particles typically move. In such models, as well as in swimming experiments, isolated individuals are free to move at some fixed speed. During swarming, on the other hand, the reason for the lack of motion of isolated individuals (or cells at low densities) is unclear. It is assumed that the cells are temporarily trapped in areas that are not sufficiently wet [Ariel 2019]. At very high densities cells cannot move efficiently due to confinement [Wensink 2012; Ariel 2018], suggesting an additional *jammed* phase J.

At intermediate concentrations, our analysis revealed three phases of motile bacteria with distinct dynamical characteristics: (i) A *swarm* phase S at small aspect ratios, in which cells move and flow efficiently, (ii) a low-density phase of long cells, which consists of *small moving clusters* SC, and (iii) a high density phase of long cells, in which *large moving clusters* LC of the size of the system cause large (but finite) fluctuations in time. We first focus on the transition between short and long cells. Note that given our relatively small number of possible aspect ratios, the precise location of the transition line and its properties (first or second, critical exponents, etc.) cannot be determined. Nonetheless, several key differences between short and long cells are apparent: At small aspect ratios (S phase), the swarm is characterized by a unimodal spatial distribution of densities (Figs. 2C-D, S4) and the velocities exhibit a Gaussian distribution (kurtosis close to 3, Fig. S5). In contrast, at large aspect ratios (SC and LC phases), the swarm is segregated into two populations, corresponding to low-density and high-density regions (Figs. 2E-F). The proportion of each population changes with the mean area coverage. The distribution of velocities exhibits very large kurtosis (indicating a heavy-tailed distribution) at the large aspect ratios (Fig. S5). Another key statistic, which has been theoretically shown to describe different regimes of collective dynamics, is the distribution of cluster sizes (DCS) [Peruani 2006]: In the S phase (Fig. 3A-B), the DCS is practically a power-law with an exponential cut-off. In contrast, in the SC/LC phases, the DCS at low densities (a power-law with a cut-off) is different than at higher ones, where large clusters emerge whose sizes are comparable with the observed system size (Fig. 3C-D). Lastly, it has been shown that WT swarming bacteria are super-diffusive, with trajectories that are consistent with Lévy walks [Ariel 2015; Ariel 2017]. Figure 2B shows that trajectories are super-diffusive at all aspect-ratios and concentrations. However, the associated characteristic exponent is varying. While the exponent in the S phase is approximately constant (1.6-1.7), it is clearly decreasing in the SC/LC phases. Therefore, the super-diffusive property of long cells degrades at high densities, indicating slower mixing and spreading.

At large aspect ratios, the phase diagram is divided into two distinct regions (Fig. 1B). The transition (as a function of density) is pronounced in the spatial correlation functions (Figs. 4A-D), both in cell directional alignment $\lambda_\parallel$, and in the velocity (direction of motion) $\lambda_v$. At low densities, the correlation lengths grow sharply with density, exhibiting small fluctuations (<5%). However, at densities higher than 0.3, the correlation lengths are practically constant, with large fluctuations between samples (up to 55%). The transition region is narrow, suggesting a critical phenomenon. Further examination reveals that the jump in the standard deviation of measurements is mostly due to density fluctuations in time (Fig. 4E). The increase in temporal fluctuations correlates with the occurrence of giant number fluctuations which, however, here only indicate the occurrence of large moving clusters, similar to previous findings in bacterial systems [Zhang 2010, Peruani2012] and in contrast to experimental reports and theoretical predictions for e.g., active nematic phases that are assumed to be homogeneous on the large scale [Narayan 2007; Ramaswamy 2010]. A time series analysis reveals a sharp increase in the Hurst exponent, which quantifies the roughness of temporal fluctuations, indicating that the density varies sharply in time (Figs. 4F, S6). For small aspect ratios, the Hurst exponent is around 0.5 at all densities, as expected for finitely correlated series.

As sketched in the introduction, our results reveal some successful theoretical predictions upon comparison to earlier work on simulation and analysis of collective behavior of self-propelled rods, but mostly in the large aspect ratio regime. Several phases that have been observed in simulations, such as the bio-nematic and laning phases [Wensink 2012; Abkenar 2013], were not realized. Other predictions, such as the emergence of bimodal cluster-size distributions [Fig. 3, Peruani 2006], motility induced spatial segregation into low and high-density regions [Mishra 2006], large number fluctuations [Figs. S7, Ramaswamy 2010] long-tailed auto-correlations functions [Figs. S8-S9, Ramaswamy 2010] and meso-scale turbulence [Wensink 2012], have been inferred from SPP and continuous models and are found, to some extent, in the present experiments. See SI Text 2 for a detailed discussion. Still, many of the prominent features of the swarm dynamics, including the non-trivial Hurst exponent (marking the SC-LC transition) and the lack of phase changes at small aspect ratios cannot be explained by current theories. Overall, our results suggest that in the S phase, bacteria do not behave like dry self-propelled rods. In this case, it is likely that hydrodynamic interactions within a self-generated lubrication layer suppresses clustering.

*B. subtilis* is a representative of a group of bacteria called 'temperate-swarmers' such as *Serratia marcescens*, *Pseudomonas*, S*almonella* and *Escherichia coli* that swarm at similar conditions [Partridge 2013]. Thus, we expect the phase diagram generated here to be qualitatively similar. Other bacteria, termed 'robust swarmers', such as *Proteus mirabilis* and *Vibrio parahaemolyticus,* are capable of migrating atop harder surfaces [Tuson 2013]. These cells are typically longer (~20 $\mu$m) and may show a truncated diagram eliminating the S phase [Little 2019]. In addition, such cells have a "life cycle" of repeated elongation, migration, division and sessility, which implies different biological functions. Moreover, the difference between phase states may be a critical determinant that differentiates moderate from robust swarmers, and thus, the kinds of surface hardness a bacterium can traverse.

From a biological perspective, bacterial swarming is a natural state, i.e., cells appear to enter a swarming state when introduced to a surface. This suggests that the changes in cells prior to the onset of swarming may be advantageous to the colony's survival. The phase diagram discussed above describes the range of possible dynamical regimes for the swarm, highlighting the subtle interplay between the physical and biological characteristics of the swarm. We find that under standard conditions bacteria inhabit a region of phase space in which the swarm dynamics is highly robust and insensitive to fluctuations. In this regime, bacteria do not cluster and do not form an orientational order that would bias the bacterial flow towards a particular direction. Such a bias reduces the assumed biological function for swarming, which is rapid isotropic expansion (given no external directional cues). In addition, the super-diffusive property of trajectories does not deteriorate at high densities. These conditions are pivotal for rapid spreading and mixing of bacteria within the swarm, which may be crucial for efficient growth and colony expansion.

From an active matter perspective, our experiments provide the first quantitative evidence of sharp phase transitions in bacterial collectives. Moreover, we are able to determine which interaction between self-propelled motile cells (steric or hydrodynamic) is dominant. In particular, we find that short and long cells exhibit distinct swarming regimes. Therefore, the phase diagram, Fig. 1B, and the subsequent detailed statistical analysis of key dynamic quantities provide a rich data set against which future models for swarming and swimming bacteria with competing alignment and hydrodynamic interactions can be calibrated or tested.

**Materials and Methods**

Growth protocol and observation

*Bacillus subtilis* is a Gram positive, rod-shaped, flagellated bacterial species, used as a model system in many quantitative swarming experiments [Kearns 2003]. Four different variants of *B. subtilis* 3610 were tested, all with the same width (~0.8 µm) but varying lengths. The cells were grown on agar plates; monolayer swarming colonies were obtained by growing the colonies on 25 g/l LB (Luria Bertani) and 0.5% agar (Difco). Plates were filled with 20 ml of molten agar and aged for 24 h in the lab (20ºC and 45% RH) prior to inoculation. The cells were incubated at 30ºC and 95% RH for about 5 h. *B. subtilis* is normally kept at -80ºC in 50% glycerol stocks and grown overnight in LB broth at 30ºC and shaking (200 RPM) prior to plate inoculation (5-µl at the center of each plate; $OD_{650}=1$, corresponding to approximately $10^7$ cells/ml).

All mutants were obtained from the same lab (Daniel B. Kearns, Indiana) [Ilkanaiv 2017]. Figure S2 (table) lists the strain name and the mean aspect ratio with the standard deviation. The mean was obtained from 500 randomly chosen cells *in the active swarm*. In most cases the large variety of cell-lengths in a specific sample is due to proliferation and cell division thus the mean cell length does not have a Gaussian distribution (size is limited by the length of a single cell and the length of twice its size).

An optical microscope (Zeiss Axio Imager Z2; 10×, 20×, 40× and 63× LD-Phase contrast lenses), equipped with a camera (GX 1050, Allied Vision Technologies) was used to capture the microscopic motion at 100 frames/sec and 1024×1024 pixels. For each aspect ratio, at least 30 independent plates were created.

Data analysis

In each experimental plate, individual cells were identified using a custom tracking software implemented in Matlab. The number density at each frame was estimated by counting the number of cells. Surface coverage was estimated by measuring the area of a threshold filter applied to either the pixel intensity or the local entropy of the image.

Snapshots were binned according to densities with a width of 0.02 surface fraction. In some of the figures, results with very sparse or very dense bins (e.g., 0.10 or 0.80) are not shown due to insufficient data.

The tracking algorithm receives as input a video of bacteria given as a sequence of frame images. It outputs the trajectories of the bacteria in the video. The algorithm is separated into two conceptual parts: Image analysis and motion reconstruction. In the image analysis step, following standard filtering and sharpening preprocessing, the location and orientation of every bacterium in each frame is obtained by thresholding an intensity histogram, separating cells from their background. Cells that overlap or are too close to be distinguished are separated using two specialized developed algorithms: The first guesses the orientation of the bacteria using axis matching. The second applies a skeleton cutting algorithm as suggested in [Liu 2012]. In the motion reconstruction step, the entire set of identified cells is used to fit a puzzle into continuous trajectories.

Measured statistics:

The dynamical properties of single cells and the swarm were quantified using several measurements and observables, which were calculated from the binned trajectory data.

- Speed: Following standard smoothing using Matlab's malowess command, speed was obtained by calculating the displacement between adjacent frames.

- Mean Squared Displacement (MSD) exponent of trajectories. Denoting the trajectory of cell $i$ by $\mathbf{r}_i(t) \in \mathbb{R}^2$, the MSD is defined as

$$\mathrm{MSD}(t) = \left\langle \left\| \mathbf{r}_i(s+t) - \mathbf{r}_i(s) \right\|^2 \right\rangle_{s,i},$$

where $\|\cdot\|$ is the Euclidean norm and $\langle \cdot \rangle_{s,i}$ denotes averaging over all times $s$ and particles $i$. Assuming that for sufficiently large $t$, $\mathrm{MSD}(t) \sim At^\gamma$, the exponent $\gamma$ is obtained using linear regression on a log-log scale.

- The spatial distribution of densities at a given average density $\rho$ was computed as follows. We collect all frames with a given aspect ratio and with overall density in the range $[\rho - 0.01, \rho + 0.01]$. Each frame is divided into $10 \times 10$ subsections. We compute the standard deviation, the kurtosis (centered, scaled fourth moment) and the histogram of densities within all subsections.

- Clustering: Two bacteria appearing in the same frame are considered neighbors if their distance between their centers is less than the mean bacterium length (which depends on the aspect ratio) and their relative speed is less than 20% the average speed in the corresponding density. Clusters are defined as connected components of the graph obtained using the above definition.

- Spatial correlation functions: For a given distance $r$, define the set $A_f(r)$ as the pairs of all cells in a frame $f$ whose centers are separated by a distance $(r-\epsilon, r+\epsilon)$, $A_f(r) = \{(i,j) \,|\, \|\mathbf{r}_i - \mathbf{r}_j\| \in (r-\epsilon, r+\epsilon)\}$ for some $0 < \epsilon \ll r$. The spatial velocity correlation is then given by

$$C_v(r) = Z^{-1} \left\langle \frac{1}{|A_f(r)|} \sum_{i,j \in A_r} \mathbf{v}_i^f \cdot \mathbf{v}_j^f \right\rangle_f,$$

where $\mathbf{v}_i^f$ is the velocity of the $i$'th cell in frame $f$, $|A_f(r)|$ is the number of elements in $A_f(r)$, $\langle \cdot \rangle_f$ denotes averaging over all frames $f$ frames with a given aspect ratio and with overall density in the range $[\rho - 0.01, \rho + 0.01]$ and $Z$ is a normalization constant such that $C_v(0) = 1$. Similarly, the angle correlation (cell orientration) is given by

$$C_\theta = Z^{-1} \left\langle \frac{1}{|A_f(r)|} \sum_{(i,j) \in A_f(r)} \cos(2(\theta_i - \theta_j)) \right\rangle_f,$$

where $\theta_i^f$ is the orientation of the $i$'th cell in frame $f$. Figure 4 shows both the correlation length obtained by averaging over all experiments in the density range (4A, B) or the standard deviation, normalized by the average (4C, D).

- Temporal auto-correlation functions: Auto-correlation functions are defined as $C(t) = Z^{-1} \langle \mathbf{r}(s+t) \cdot \mathbf{r}(s) \rangle_{s,f}$, where $Z$ is a normalization constant such that $C(0) = 1$ and $\langle \cdot \rangle_{s,f}$ denotes averaging with respect to all trajectories in frames at a given density range and times $s$ along a trajectory. We study three auto-correlation functions in different vector fields $\mathbf{r}(t)$: the instantaneous velocity, normalized instantaneous velocity (direction of movement) and cell orientation.

- Hurst exponent $H$: Given a time series, the Hurst exponent quantifies the roughness of fluctuations in the series. For a time series $X_1, X_2, \ldots$, define $h(n) = E[R(n)/S(n)]$, where $R(n) = \max\{Z_1, \ldots, Z_n\} - \min\{Z_1, \ldots, Z_n\}$ is the range of the cumulative (centered) first $n$ observations, $Z(n) = \sum_{i=1}^{n}(X_i - E[X_i])$, and $S(n)$ is the standard deviation obtained in the first $n$ observations. Assuming that asymptotically, for large $n$, $h(n) \sim Bn^H$ defines the Hurst exponent $H$. When data is sparse, the Hurst exponent can be approximated using the following method, which was used here: Let $w(n) = \langle \text{std}\{X_k, \ldots, X_{k+n-1}\}\rangle_k$, i.e., the average standard deviation of $n$ sequential observations. Then, for large $n$, $w(n) \sim Bn^H$. Figure 4F was generated using the density obtained in ×40 magnification.
- Large number fluctuations: Cells with particular aspect ratio and density were partitioned into 1-30 bins in each dimension. The variance among all bins was calculated.

**Acknowledgments:**

Partial support from The Israel Science Foundation's Grant 373/16 and the Deutsche Forschungsgemeinschaft (The German Research Foundation DFG) Grant No. HE5995/3-1 and Grant No. BA1222/7-1 are thankfully acknowledged.

**Fig. 1.**

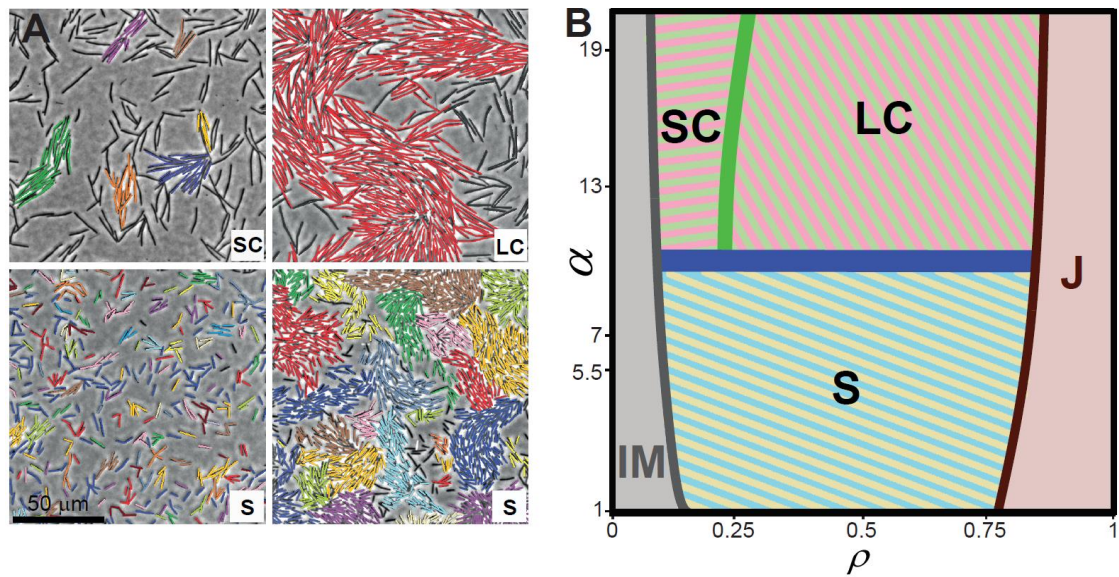

**Figure 1: A phase diagram for bacterial swarming. A**, Snapshots of monolayer bacterial swarms with different aspect ratios and densities, representative of the motile phases. Colors represent moving clusters (gray or non-labeled cells are not moving). For short cells (S phase), almost all cells are moving coherently. For long cells, spatial ordering into local (SC) or global (LC) clusters is apparent. **B**, The inferred phase diagram, showing five distinct phases: Immotile (IM), swarming (S), small clusters (SC), large clusters (LC) and jammed (J).

**Fig. 2.**

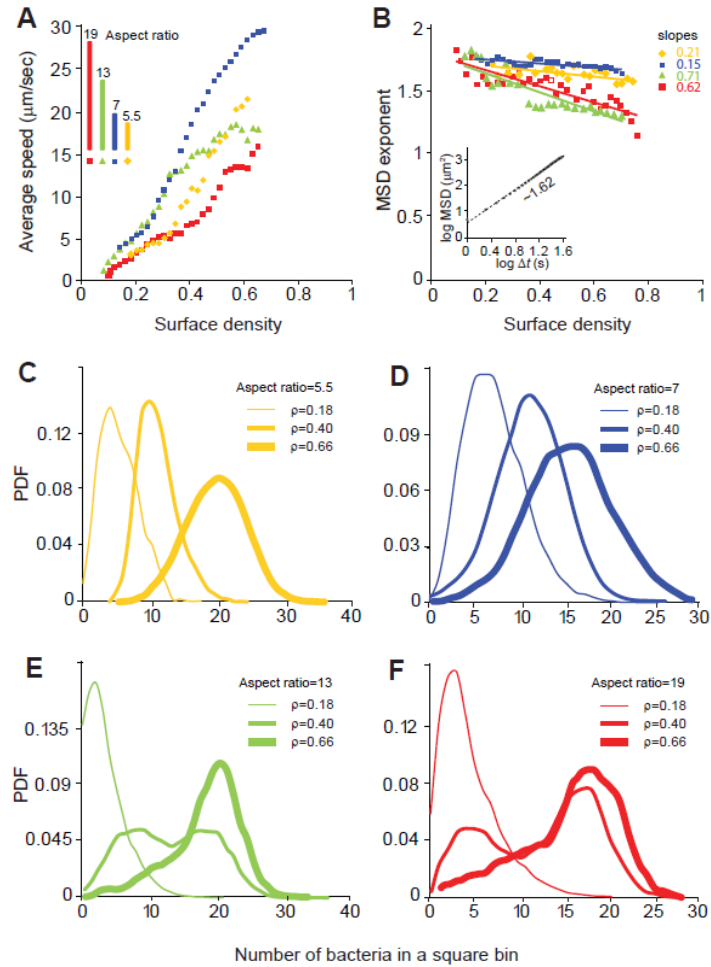

**Figure 2: Short vs. long cells. A**, The denser the cells, the faster they move. The average speed increases with density, but is non-monotonic in aspect ratio. Wild-type cells are the fastest. **B**, The Mean-Squared-Displacement exponent shows that cell-trajectories are super-diffusive. However, the exponent of long cells (SC/LC phases) decreases rapidly with density, showing that their super-diffusive property degrades at high densities. Inset shows an example of the log-log plot from which we have generated the slope values (in this example the hollow red-square data point). **C-F**, The distribution of densities among a 10×10 partition of the viewing area for different (average) concentrations. **C-D**, At small aspect ratios (S phase), the swarm is characterized by a unimodal spatial distribution, with the mode increasing with mean surface coverage. **E-F**, At large aspect ratios (SC and LC phases), the swarm is segregated into two populations. The proportion of each population changes with the mean area coverage.

**Fig. 3.**

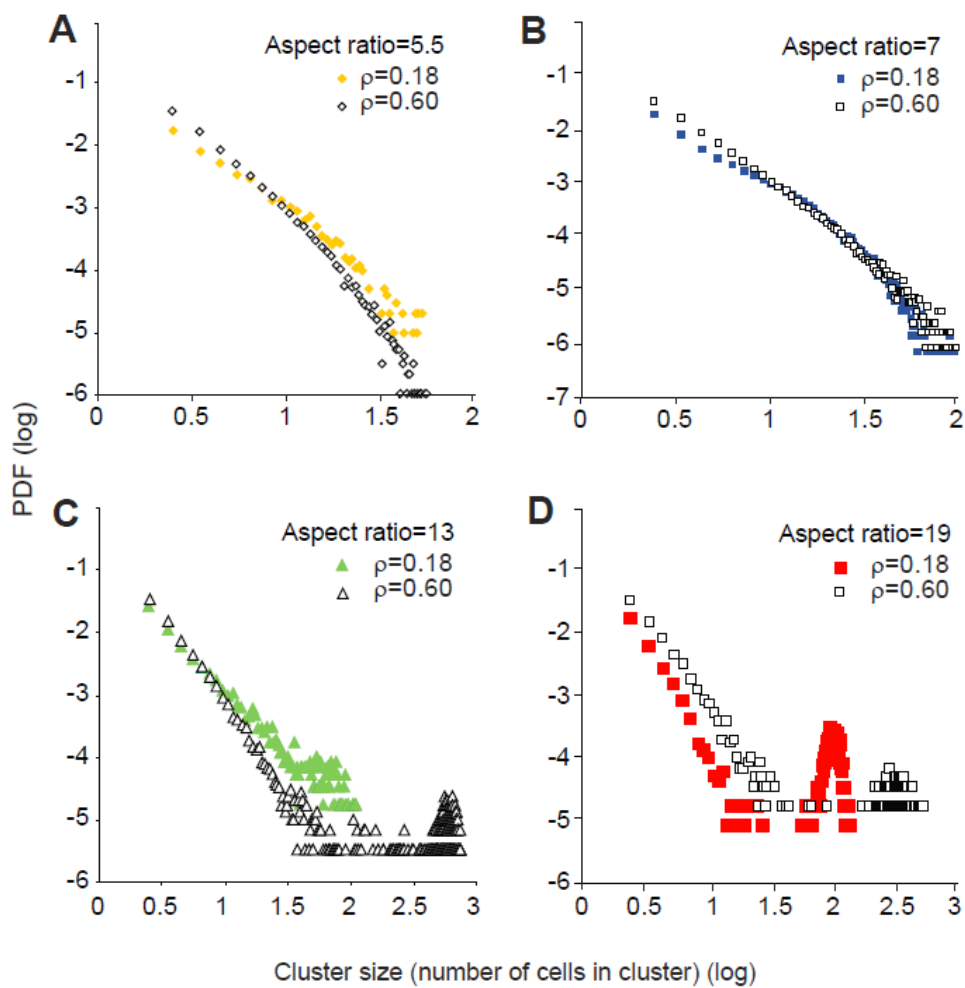

**Figure 3: Distribution of cluster sizes. A-B**, For short cells (S phase), the distribution of cluster sizes is approximately a power-law with an exponential cutoff. **C-D**, For long cells (SC/LC phases), the distribution is more complex: at low densities, it is a power-law. However, at higher densities e.g., $\rho=0.6$, large clusters emerge whose sizes are comparable with the system size.

**Fig. 4.**

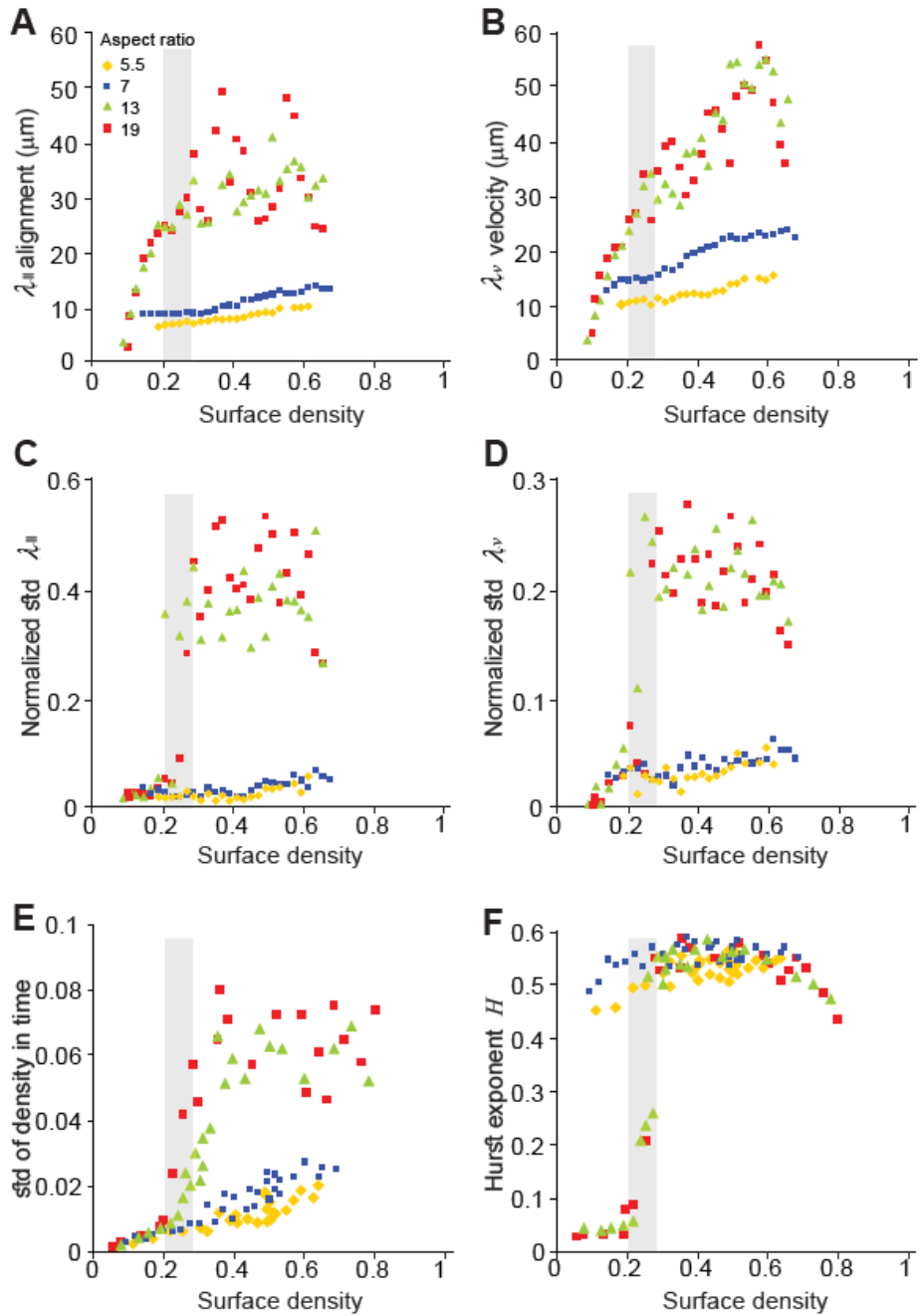

**Figure 4: Characterizing the SC-LC transition. A**-**D**, Correlation length in cell direction (**A**, **C**) and velocities (**B**, **D**). **A** and **B** depict averages and **C** and **D** standard deviations – showing a sharp increase within a narrow region of densities. **E**, The standard deviation of temporal fluctuations in density suggests large number fluctuations in time. **F**, The Hurst exponent, which indicates the roughness of a time series, shows a sharp jump in the SC/LC transition region. The gray rectangle represents the transition region.

**Supplementary Text 1**

Control experiments

In order to verify that the differences in the swarming dynamics statistics are only due to aspect ratio, several control tests were performed. This is important because the 3 strains were generated in different ways, genetically affecting specific genes to form short or long cells. These tests guaranty that on the single cell level, cell motility was not affected. See Figs. S2 and S3.

**Supplementary Text 2**

Swarming bacteria and self-propelled rods

A popular model to account for the basic physics of interacting self-propelled agents such as swarming bacteria is the self-propelled rods (SPR) with steric interactions [Peruani 2006; Wensink 2012; Abkenar 2013; Weitz 2015; Shi 2018, Großmann 2019, Bär 2019]. The experiments presented here do have features predicted in simulations of self-propelled rods, such as the transition from an exponentially decay cluster size distribution (CSD) in the swarming phase to a clustering phase with a bimodal CSD as shown in Fig. 3. Such a behavior has already been reported for a mutant myxobacterial species [Peruani 2012]. The transition to glassy resp. jammed dynamics at large densities is also a feature often captured by SPR models, see e. g. [Abkenar2013, Weitz2015]. For the simulations of SPR with steric repulsion cited above, the product of the aspect ratio and density has often turned out to be a crucial control parameter for the onset of collectivity as well as for the emergence of spatial patterns such as polar cluster reflecting a basic property of Onsager´s famous theory for the isotropic-nematic phase transition in colloidal rods [Onsager 1949]. This property is in contrast to the observation presented here that the transition from the swarming phase to the clustering phases is only weakly dependent on the density and shifted also to larger aspect ratios in comparison to SPR simulation.

In addition, the bacteria studied here also require proximity to sufficiently many other bacteria in order to move on the agar surface, whereas individual SPR move at a prescribed typical speed independent of the local bacterial density. This strong nonlinear dependence of the velocity on the density should also be incorporated in a future model in order to achieve a quantitative description of the experiments in this work. A direct inspection of the snapshot in the clustering phase shows a qualitative difference to the images recorded for the myxobacteria mutants that exhibit typical SPR dynamics [Peruani 2012] as well as to images of SP rods [Abkenar 2013; Weitz 2015]. The experiments above also indicate that the bacteria studied here need a liquid surrounding for their flagellar motion and therefore are subject to hydrodynamic interactions in addition to short range steric repulsion encoded in the SPR models. The weakening of the tendency to form clusters has recently been reproduced also by simulations of SPR with hydrodynamic interactions embedded in a fluid [Theers 2018].

Our results on number fluctuations presented in Fig. S8 also strongly deviate from the results for different models of self-propelled rods and related experiments with gliding bacteria [Ginelli 2010, Peruani 2012, Nishiguchi 2017], which all exhibited giant number fluctuations (GNF) with an exponent of ~0.8. Here, most experiments, in particular for the wild-type and short bacteria, show less pronounced GNF with exponent in the range of 0.5-0.6. The large values seen for SP rods without hydrodynamic interaction are approached only for very elongated bacteria (with aspect ratio > 10) at intermediate densities and far away from the transition towards the immotile and jammed phases.

**SI references**

**Fig. S1.**

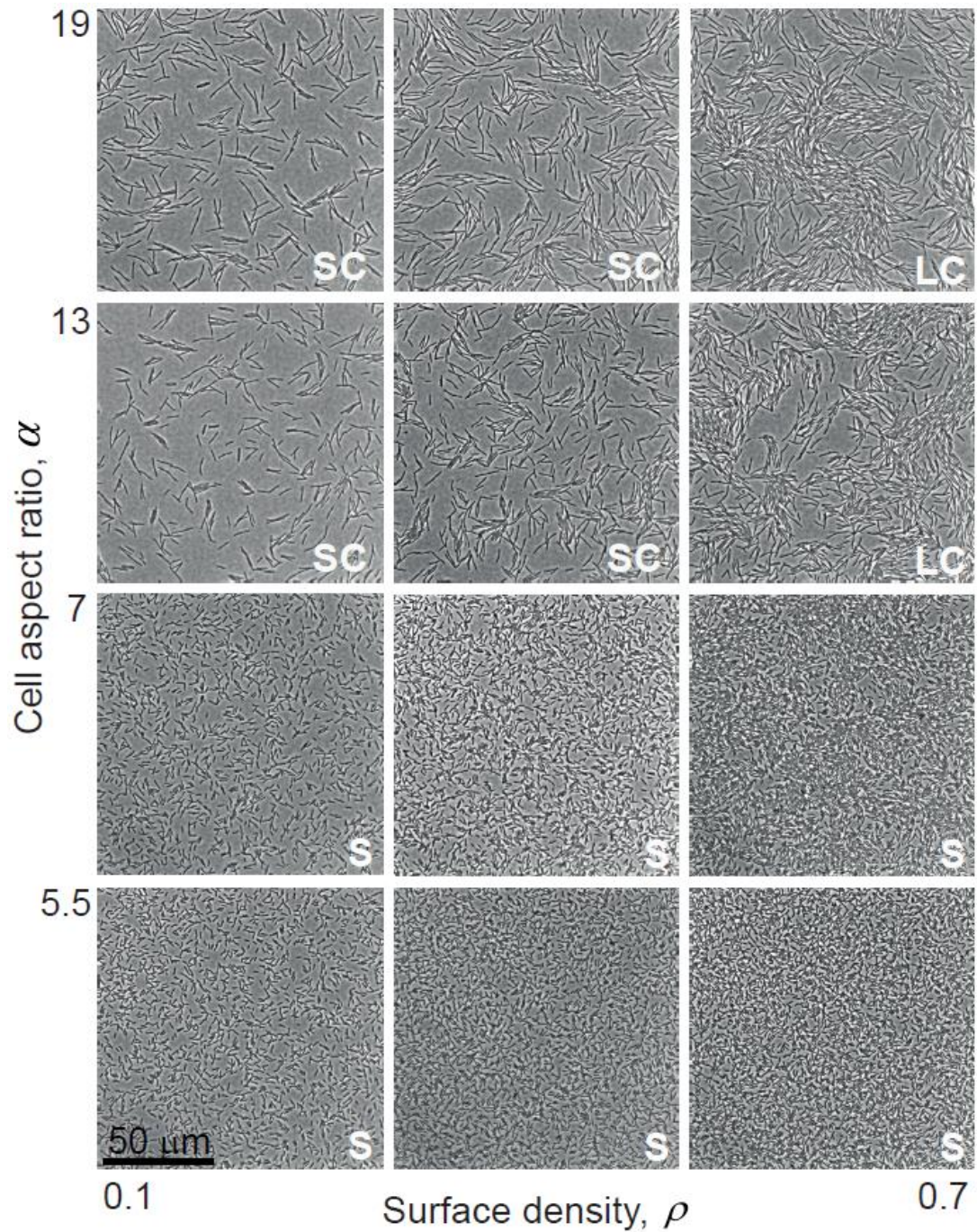

**Figure S1: Snapshots of monolayer bacterial swarms with different aspect ratios and densities.** Similar to Fig. 1A but for a much larger field of view.

**Fig. S2.**

| Strain | Aspect ratio in swarm | Mutation |
|---|---|---|
| DS1470 | 5.5±1.5 | swrA::tet amyE::Physpank-swrA spec |
| WT | 7.0±2.2 | - |
| DS3774 | 13.0±4.2 | minD::TnYLB kan |
| DS858 | 19.0±5.6 | minJ::tet |

**Figure S2: Strains data - Table.** A summary of the *B. subtilis* 3610 strains used in this work.

**Fig. S3.**

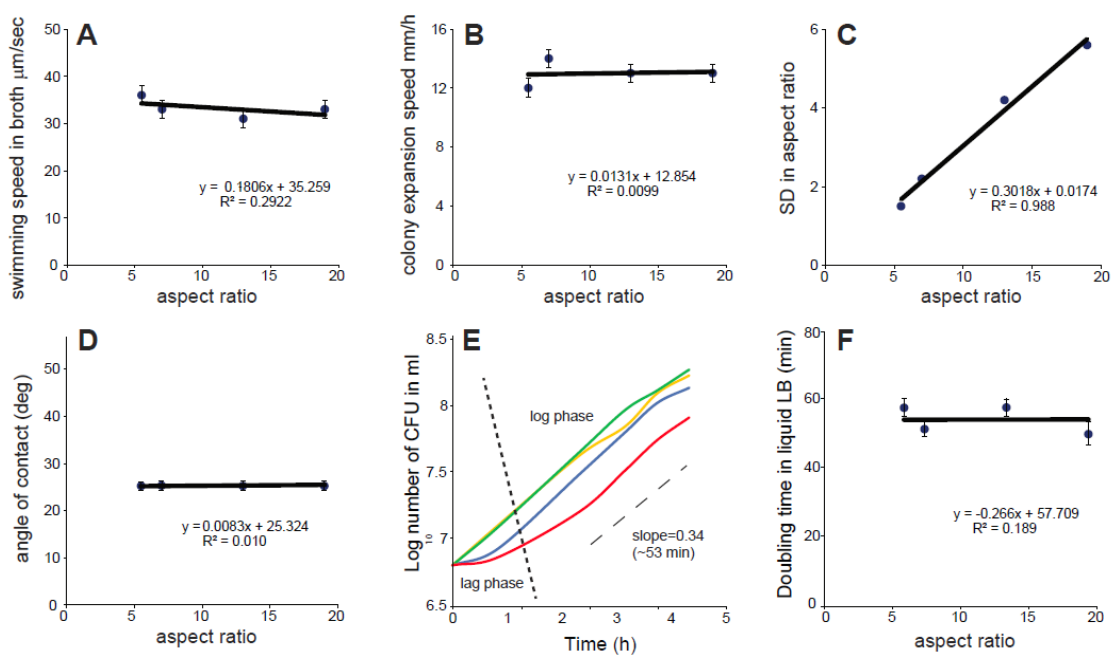

**Figure S3: Motility control experiments. A**, The swimming ability of the bacteria in sparse ($10^4$ cells/ml) liquid suspensions, where interactions between cells are negligible and there is no interaction with a surface, is the same for all strains. Both wild-type and the 3 mutants swam in the well-known 'run-and-tumble' mode with approximately similar speeds (~33 μm/s) during 'runs' (mind that the cells in this experiment do not grow to the same aspect ratio as in the swarm experiment). **B**, The ability of all strains to colonize LB plates is fairly constant. All strains showed a macroscopic swarm pattern and covered a standard Petri-dish (8.8 cm) within a few hours, moving the colony front at approximately the same speed (~13 mm/h). **C**, Here we show a linear trend in the standard deviation of cell aspect ratio for each strain as a function of mean cell aspect ratio. This indicates that relatively small, or large, deviations in cell length within a strain are not affecting the collective dynamics. **D**, Drops of overnight cultures were deposited on flat smooth plastic surfaces (10 μl on the inner part of the tap of a Petri-dish). The contact angles indicate the amount of surfactants secreted by cells. This angle (~25°) was nearly identical for all strains (note that the surfactant experiments were performed on bacteria taken from swimming cells where swarming motility is not expressed; the cells in this experiment do not grow to the same aspect ratio as in the swarm experiment). **E-F**, Growth curves for the four strains were obtained. Each strain

was grown overnight from a plate stock (with the appropriate antibiotics) in a separate 50 ml tube in fresh LB broth (with no antibiotics) at 30ºC and shaking (200 RPM) (10 ml in each tube). The overnight cultures were diluted (1/100 fold) and regrown at same conditions. The number of CFU grown on 1.5% agar LB plates from these suspensions (after appropriate culture dilution) for each strain was counted; this was repeated during few hours from first dilution. The experiments were done few times so that the optical density was adjusted in all strains to yield same number of CFU in the starting culture. A summary of these measurements shows that the doubling time is similar in all strains (approximately 53 min) with very small variations in the duration of the lag phase. Overall, these results indicated that all mutants have similar swimming and swarming capabilities with no apparent or immediate motility defects.

**Fig. S4.**

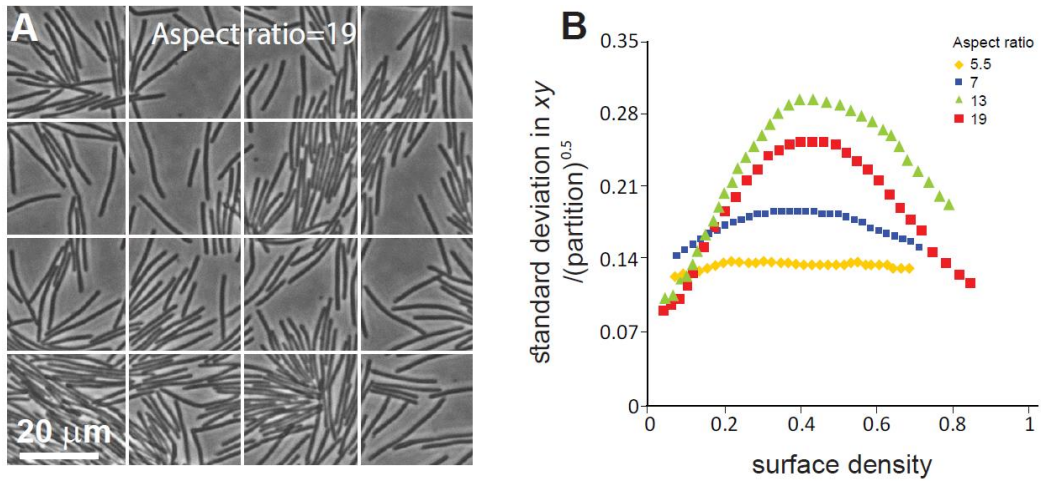

**Figure S4: Spatial distribution. A,** In order to estimate the spatial distribution of cells, all frames corresponding to a given density were partitioned into 10×10 bins (for illustration purposes, we only show a part of the full viewing field). **B**, The standard deviation among bins as a function of density and aspect ratio.

**Fig. S5.**

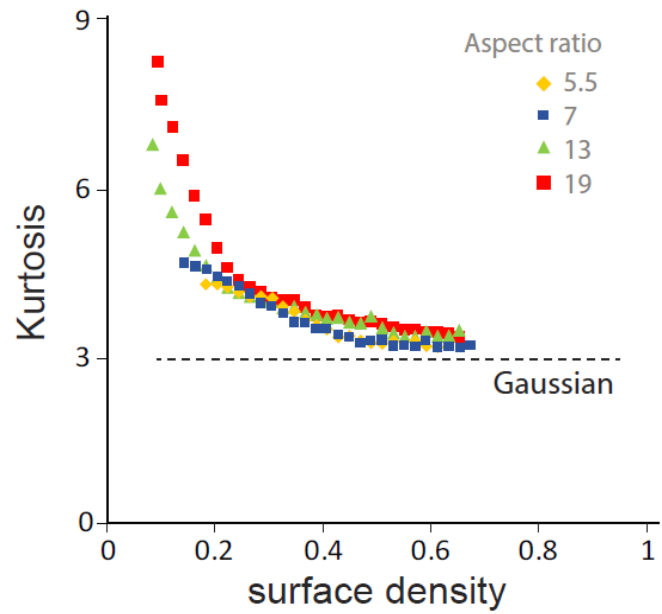

**Figure S5: Distribution of velocities.** At small densities, the kurtosis (4$^{th}$ moment) of the velocity distribution is much larger for large aspect ratios.

**Fig. S6.**

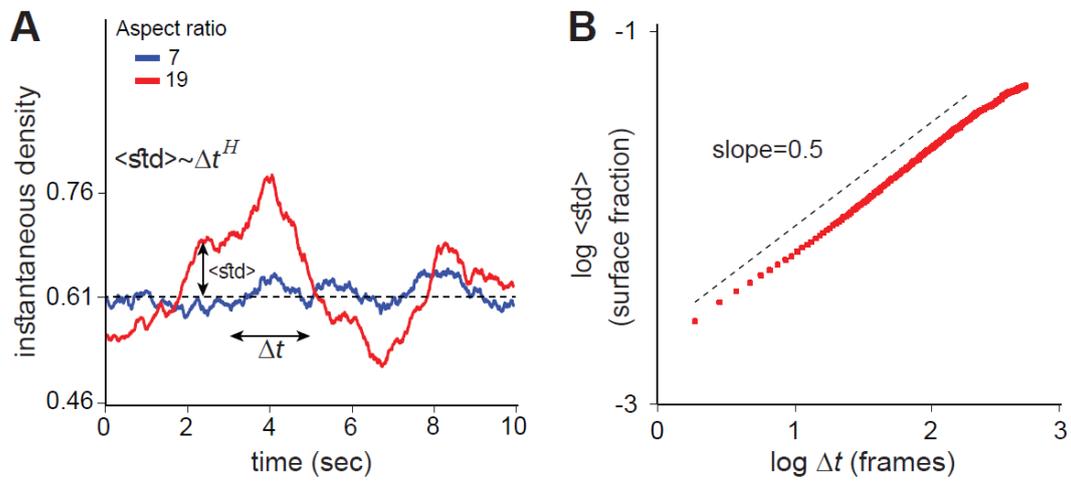

**Figure S6: Temporal fluctuations.** The Hurst exponent may be obtained for any time series or landscape, by taking the mean standard deviation <std> around the mean value for different "windows" of size $\Delta t$. If the time series (the landscape) is self-similar one will obtain a scaling of <std>$\sim \Delta t^H$ with $H$ being the Hurst exponent. $H$ Indicates the roughness of the landscape; with $H \sim 0.5$ for a regular random walk $H \sim 0$ for bounded fluctuations.

**Fig. S7.**

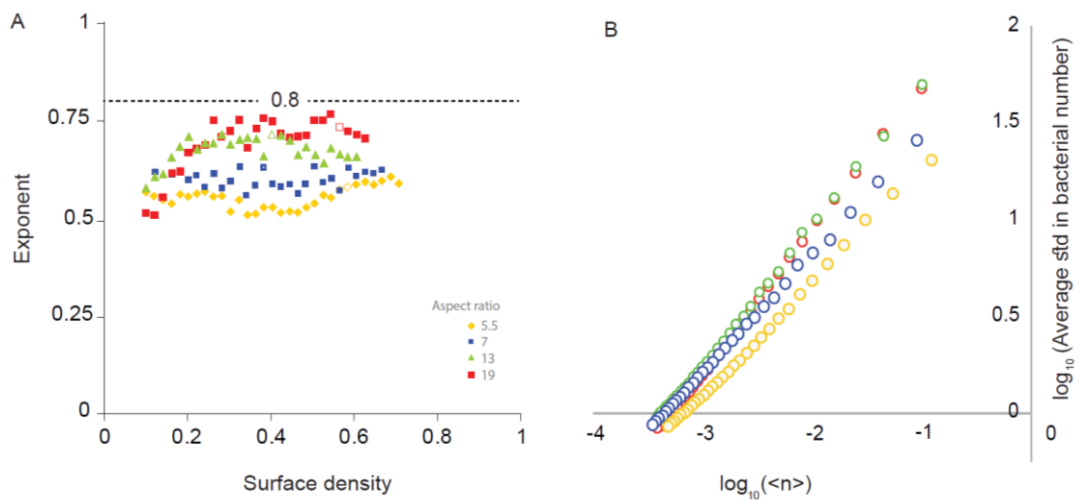

**Figure S7: Giant number fluctuations. A,** The variance in the density among sub-regions as a function of aspect ratio and average density. See the methods section for details. While giant number fluctuations are observed, the effect is not as pronounced as predicted by theory and simulations. Each point corresponds to the slope of a log-log plot showing the standard deviation in sub-domains as a function of the average number of cells (relative to maximal concentration). **B,** four example plots corresponding to the four empty points in A.

**Fig. S8.**

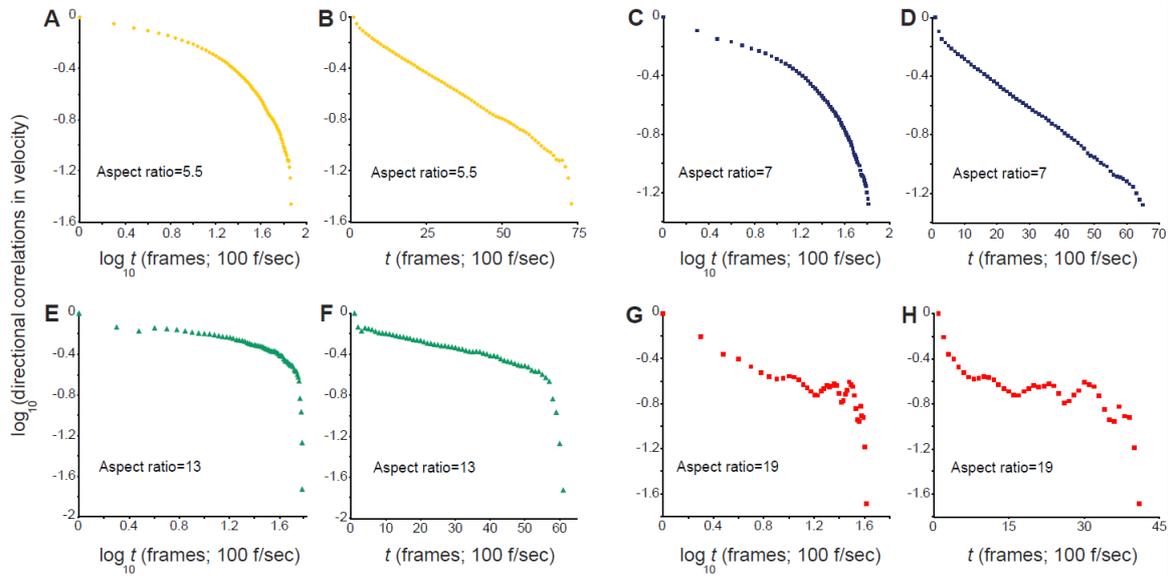

**Figure S8: Auto-correlation functions – velocity direction.** The temporal auto-correlation function in the velocity direction decays (approximately) exponentially at all densities and aspect ratios (see methods for definition). The figure depicts four examples plotted at both log-log (**A**, **C**, **E**, **G**) and semi-log (**B**, **D**, **F**, **H**) plots. **A-B**, aspect ratio=5.5, density=0.52, **C-D**, aspect ratio=7, density=0.52, **E-F**, aspect ratio=13, density=0.52, **G-H**, aspect ratio=19, density=0.52. The characteristic time scale of short cells (S phase) is about twice as small compared to long cells (SC/LC phases). The auto-correlation in the velocity behaves similarly.

**Fig. S9.**

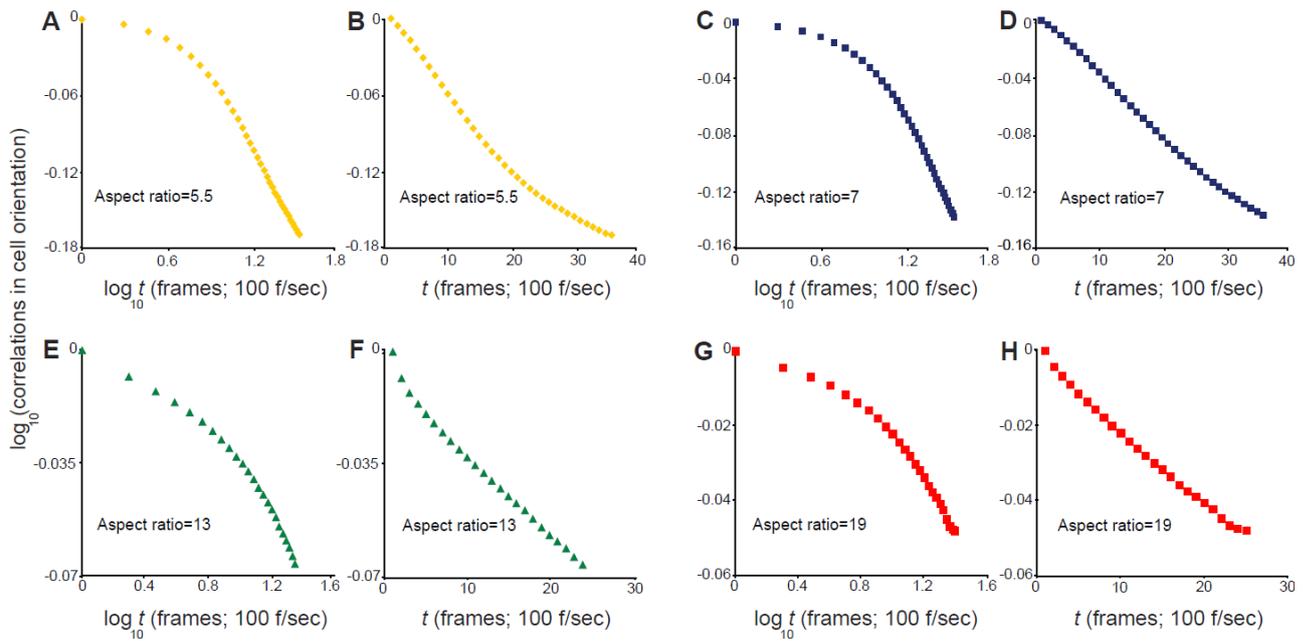

**Figure S9: Auto-correlation functions – cell orientation.** The temporal auto-correlation function in the cell orientation (see methods for definition). The figure depicts four examples plotted at both log-log (**A**, **C**, **E**, **G**) and semi-log (**B**, **D**, **F**, **H**) plots. **A-B**, aspect ratio=5.5, density=0.52, **C-D**, aspect ratio=7, density=0.52, **E-F**, aspect ratio=13, density=0.52, **G-H**, aspect ratio=19, density=0.52. The auto-correlation function of short cells (S phase) shows two regimes – at short times the decay is exponential. At long times the decay seems to resemble a power-law. However, our statistics is not sufficient to determine this with accuracy. The auto-correlation function of long cells (SC/LC phases) seems only exponential, although it is possible that the power-law behavior occurs at longer times, for which we have insufficient statistics. The characteristic time scale of short cells is about 2-3 times smaller compared to long cells.